# Ionization Aerodynamic Model of the Ionocraft

Ma Chen [a], Guo Jia [b], Tang Yuanhui [a], Lu Rong-de [b]*, Ye Bang-jiao [b]

[a] School for the Gifted Young, University of Science and Technology of China, Hefei, Anhui 230026, PR China

[b] School of Physical Science, University of Science and Technology of China, Hefei, Anhui 230026, PR China

*Corresponding author: lrd@ustc.edu.cn

**Abstract:** An ionocraft is a small instrument that generates net vertical force from electrostatic field. A model of ionization and flow is built to explain this force. The assumption of a threshold field density and the constant charge density is applied to simplify the flow model. With these assumptions, corresponding simulation results and comparison with experimental data, we have shown that this model is sufficient to explain the force on the ionocraft.

## 1. Introduction

An ionocraft, also called lifter, is an electrostatic instruments providing lifting force. Its fundamental structure is simple, as shown in fig. 1. Since its discovery in 1920s, there have been many works on this phenomena [1-4]. Among which, some theoretical models have been established and tested, such as ion wind model [5,6], ion drift model [6] and zero-point energy model [6,7]. However, there hasn't been any satisfactory explanation of the principle behind the phenomenon. Our latest work made an attempt toward this principle.

The ionocraft instrument is no more than a combination of electrostatic field and a medium of air, and the origin of force must be among them. Here we list several possible phenomenon to generate force:

(1)Ionization:

Under strong electrostatic field, some medium such as air will be ionized, so the generated positive and negative charges will move and ionize some other medium molecules. This is a complex process and is discussed in detail in gas discharge [8, 9]. These charges, if static, may generate a net electric field; and if moving, may lead to fluid effects.

(2)Fluid effects:

The medium such as air will flow under a proper source of force [10, 11], and we need to take this into consideration. Because this flow is caused by electric field alone, we can put ionization and flow together, considering the 'counterforce' by the fluid on ionocraft.

(3)Other possible origins:

There are some possible ways such as zero-point energy by Jiang Xingliu [7], which is not concerned in this issue, may be too small compared with the possible causes listed above. Another reason is that no such effects has been recorded in a vacuum except for a vague video from unconfirmed source [6].

Since polarization of air is restricted in a very small region, it is ignored in our research. We will focus on the other two effects: ionization and flow.

## 2. The ionocraft geometry model

The ionocraft can be made in many form: triangular, circular, and so on. However, every ionocraft includes a lifting unit. The ionocraft liftimg unit is shown in fig.1. It is more comprehensible in another name (asymmetrical capacitor). It has a small-scale positive electrode of thin wire, and a relatively large negative electrode of metal foil. Since it is highly symmetrical, we will only analyze 2 dimensional properties.

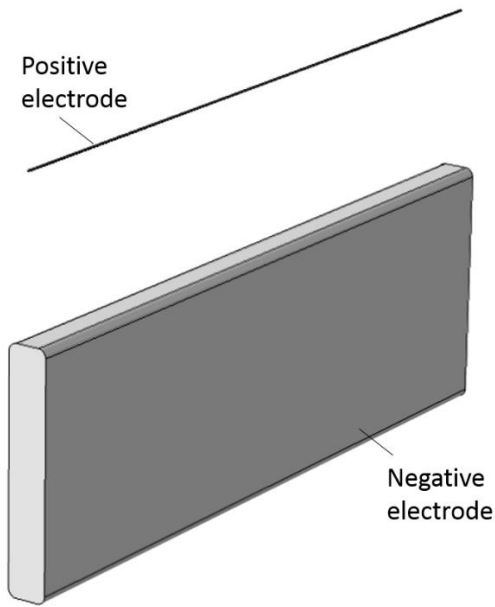

Fig. 1: The lift unit of ionocraft is more comprehensible in another name (asymmetrical capacitor). It has a small-scale positive electrode of thin wire, and a relatively large negative electrode of metal foil.

A 2D model for calculation is shown in fig.2. A small circular region of radius r1 represents positive electrode, and a larger region of corner-refined rectangle represents negative electrode. The region for calculation is the cylindrical region of radius Rin, its outer boundary will be named Rin boundary and will be mentioned in fluid calculation. All parameters in this model are listed in fig.2(d) where beta and ThV will be explained in section 4.1, and L is the length of the ionocraft on z-axis.

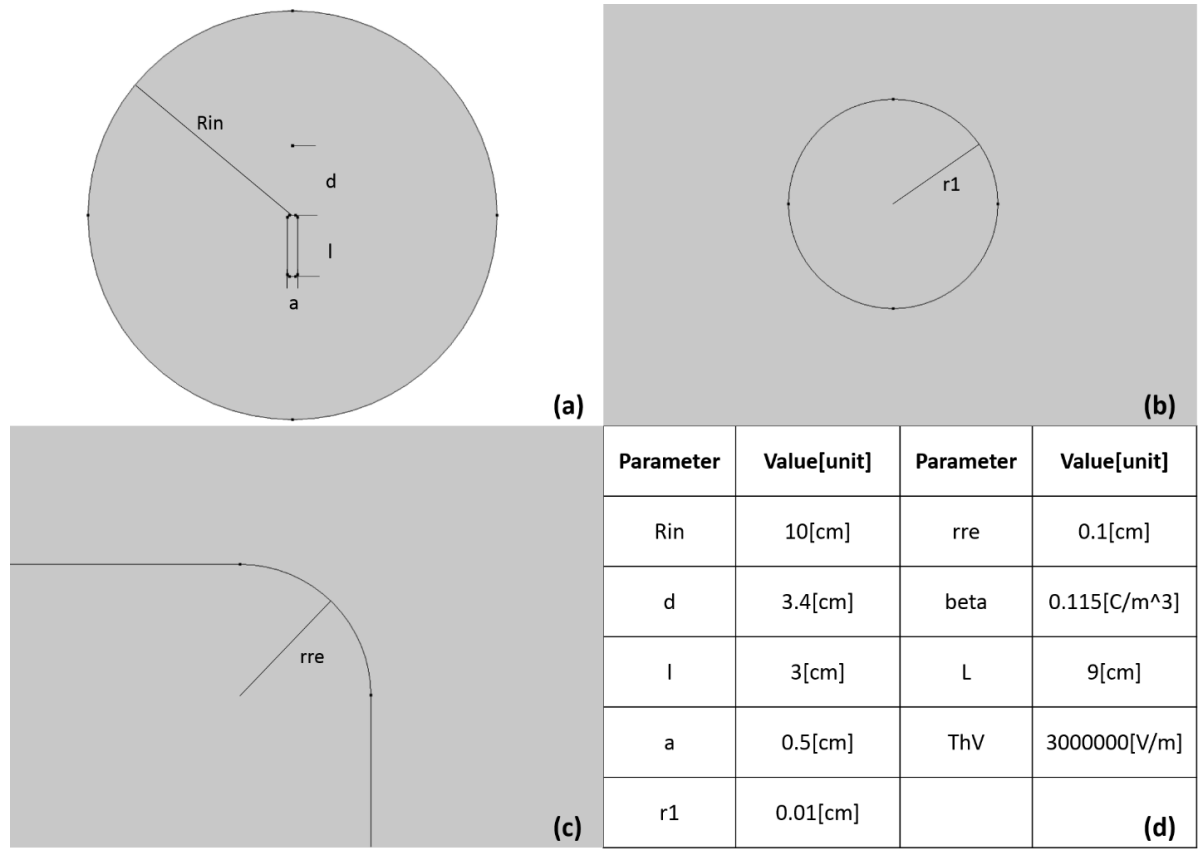

Fig. 2: The 2D geometrical model for the ionocraft. (a)Overview, it contains a tiny positive electrode and flat negative electrode, separated by a distance *d*. *l* is the vertical length of negative electrode. a is the horizontal width of negative electrode. The region for numerical analysis is indicated by the parameter Rin; (b) Closer look at positive electrode. The radius of electrode is indicated by r1; (c)Closer watch negative electrode. A refinement to corners of the rectangle with a circle of radius rre is applied to avoid singularity; (d) Values of parameters.

3. **Model establishment**

   3.1. Ionization mechanism and assumptions

   Medium under high electric field will be ionized. Here we will not begin from the mechanism of ionization, and instead, some assumptions will be made to simplify this problem.

   **Assumption 1**: in a certain region, where the electric density is larger than a threshold, the air is completely ionized, which means a certain proportion of molecules will be charged by 1 unit; out of this region, the air is completely not ionized. This threshold is represented by ThV in our model. Meanwhile, the proportion is represented by the charge density β and will be discussed in section 3.2.

   **Assumption 2**: charges uniformly and statically distribute in the ionization region, meaning volume force is simply a multiplication of $\beta$ and electric field density $\vec{E}$.

With these two assumptions, we ascribed Ionocraft to an interaction between electric field and air, described by two parameters: ThV and $\beta$, and each of them has explicit physical meaning. Compared with former research [1], we reduced the number of parameters from 7 to 2.

We should keep in mind that $\beta$ is still an averaged charge density when we assume it is a constant with space variables. Its effectiveness will be clarified by comparing with total charge density available for air in common condition. Compared with the total charge density possible of $3.90 \times 10^6 \, C \cdot m^{-3}$ if one molecule provides one unit and under 1atm and 300K, our choice of $\beta = 0.115 \, C \cdot m^{-3}$ is quite a small proportion.

3.2. Macroscopic flow mechanism and assumptions

For given ionization region, charges within it will be accelerated by electric field, and deliver their momentum by collision to regular molecules out of the ionization region [12]. This electric force, however, cannot be totally determined unless we could know the proportion of ionization. We have introduced parameter $\beta$ for charge density to represent this proportion.

We can calculate the counterforce by $F = -\int \nabla p \, d^3 x$. Here we need to notice that this force is not caused by pressure difference, for the area of horizontal projection is still small. Instead, we treat the force as a recoil of flow instead of a pressure effect. Some researches has shown that such conversion of energy from electric field to medium is not large [13, 14]. However, as will be shown in the following section, its effect is enough to explain the force on the ionocraft.

## 4. Simulation results

4.1 Simulation setup

COMSOL Multiphysics 4.4 is used in simulations. Electrostatics and turbulent flow k-$\epsilon$ module are used. Geometry model has been shown in fig.1, and built-in material air is used as medium.

In electrostatics module, a high voltage $V = 35000 \, V$ is applied on the positive electrode, and ground boundary condition is applied on the negative electrode. No boundary condition is applied on Rin boundary.

In turbulent flow k-$\epsilon$ module, a reference pressure of $1 \times 10^2 \, [atm]$ is set to avoid divergence. As explained in the last section, volume force in our problem is applied in the following form:

$$Fx = \text{if}(\text{es.normE} > \text{ThV}, \beta \cdot \text{es.Ex}, 0) \quad (2)$$

$$Fy = \text{if}(\text{es.normE} > \text{ThV}, \beta \cdot \text{es.Ey}, 0) \quad (3)$$

And the force result is shown in fig.3 (c). The red ring shows the boundary of $\text{es.normE} = \text{ThV}$, and out of it, the force goes to zero.

In addition, for the convenience of solving, we set an inlet and an outlet, as shown in fig.3 (d). Inlet is painted in blue and velocity is set to be 0, meaning we set Rin boundary as infinity here; outlet is painted in green and the pressure is set to be 1[atm].

By default setting, we use MUPUS for the computing electric field, and PARDISO for the fluid solution. Most of other parameters in calculation remain default.

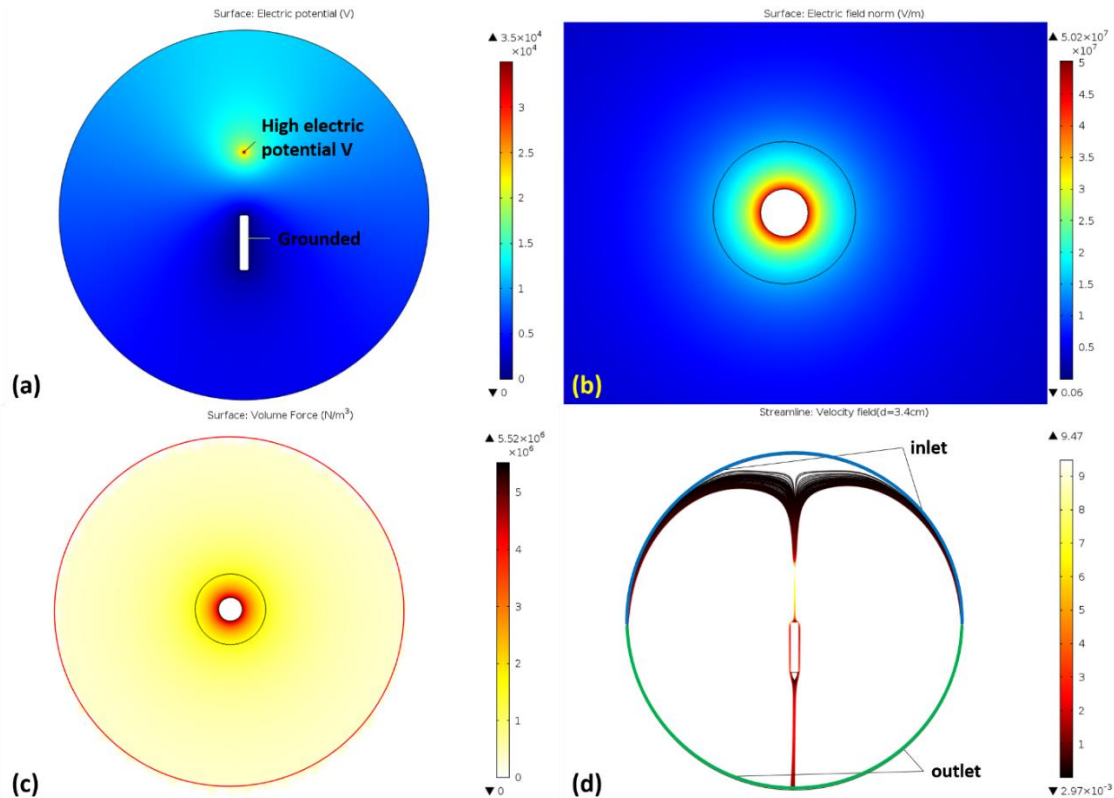

Fig. 3: (a) Electric setup and electric potential result of the ionocraft. A high voltage $V = 35000$ V is applied on the positive electrode, and the ground boundary condition is applied on the negative electrode. No boundary condition is applied on Rin boundary. (b)Field density result is highly concentrated around positive electrode. (c) Volume force setup of fluid module, threshold field is set to be $3 \times 10^6$ V·m$^{-1}$, which is the common value for air. The red ring shows the boundary where force vanishes (d) Flow module setup, Inlet is painted in blue and velocity is set to be 0, meaning we set Rin boundary as infinity here; outlet is painted in green and pressure is set to be 1[atm].

4.2 Simulation results

There are two main parts in simulation result: velocity field and d-force relation. The velocity field result shows some basic properties of this flow; while d-force relation shows this model can explain observed force.

An overlook of velocity field is shown in fig.4 (c). We can see a clear jet beginning on the positive electrode, splitting on the negative electrode, and end on Rin boundary. In addition, we may take look at the velocity magnitude $10^0 \sim 10^1$ m.s$^{-1}$. This velocity is sometimes sufficient for further application.

Most importantly, we calculated how force change with the d–distance between electrodes. As shown in section 3, we cannot calculate beta from first principle, thus we need to get its value from the experimental data. Several values of β have been tried when d = 3.4cm and finally we set $β = 0.115[C \cdot m^{-3}]$, and sweep on d with this β. Result is shown in fig. 5. We can see simulation results fitting experimental data quite well.

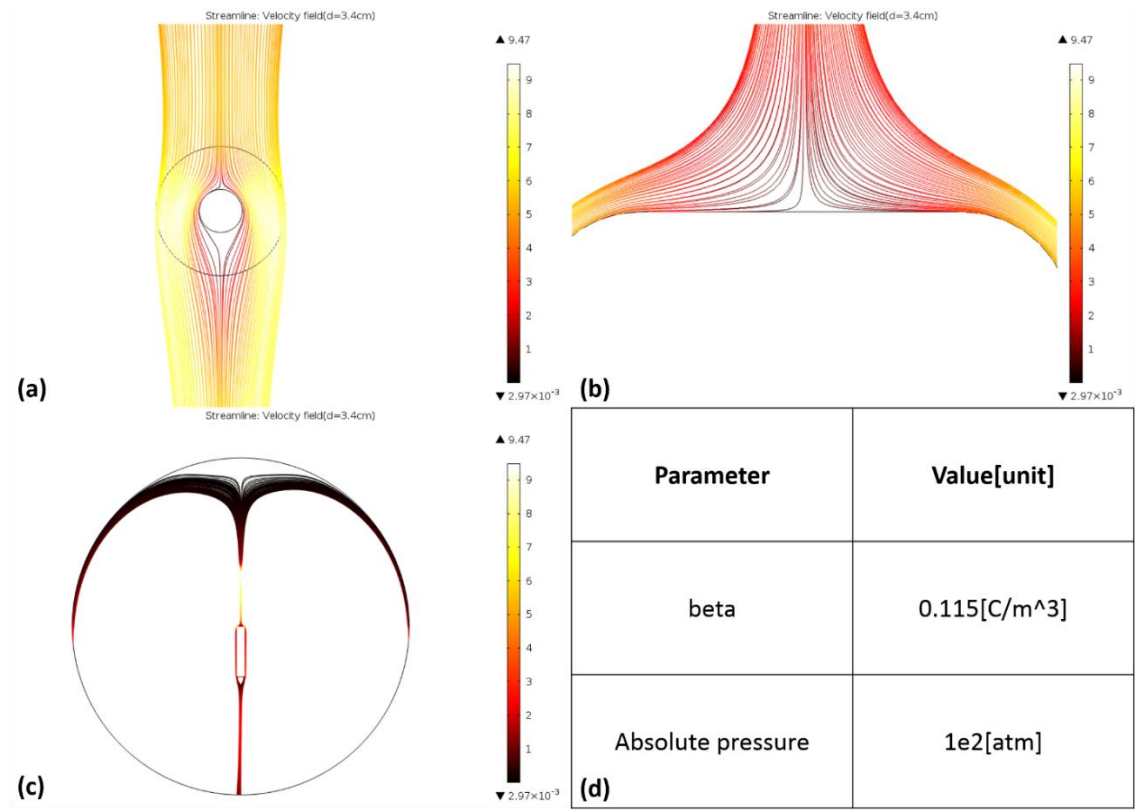

Fig. 4. (a) A detailed view of jet around positive electrode. (b) A detailed view of jet around negative electrode. (c) Overview of jet. A clear jet can be seen beginning on the positive electrode, splitting on the negative electrode, and end on Rin boundary. (d) Parameter values.

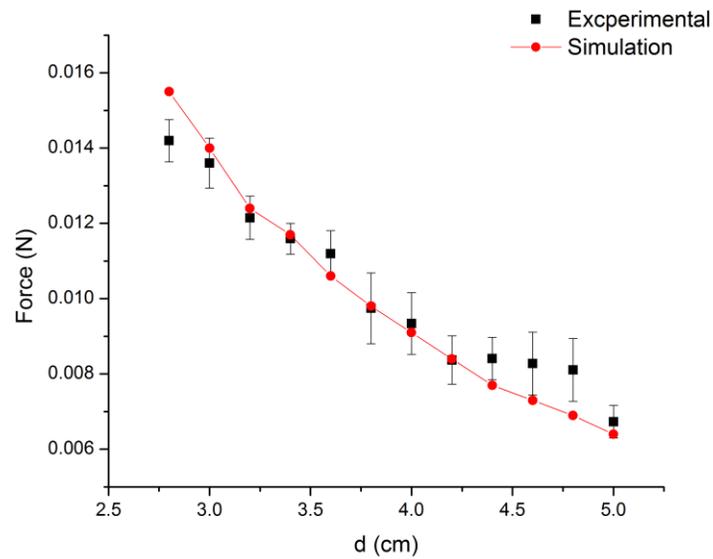

Fig. 5. Experimental data and simulation results of $\beta = 0.115 [\mathrm{C \cdot m^{-3}}]$

## 5. Conclusion

We have set up a model of ionization and flow to explain the force on the ionocraft, and shown with d-force relation that the model fits quite well with experimental data. We believe our research could confirm that the air and electric field alone would be enough to explain the ionocraft. However, there still remains some work to do a more accurate experiment needed to test our model. Other properties of the ionocraft are needed to be tested for further research. We hope our research could bring more attention on the ionocraft for more detailed research.

It also inspired us that the value β is such a small proportion of ionization that if we can make it larger, a much larger force will be reached and this instrument may one day be practical to carry things; furthermore, if we build a wing-like negative electrode, we may be able to use this flow in another way, which may be much more sufficient in the energy-force efficiency.

**Acknowledgements**

This work was supported by the National Natural Science Foundation of China (Grant No. 60974038), the Project of provincial Teaching Research in Anhui Institutions of Higher Education (No. 2012jyxm006) and the Project of the State Education Commission Teaching Research in Institutions of Higher Education (WJZW-2010-hd).